\documentclass[%
 reprint,
superscriptaddress,
 amsmath,amssymb,
 aps,
 pra,
]{revtex4-2}

\usepackage{graphicx}% Include figure files
\usepackage{dcolumn}% Align table columns on decimal point
\usepackage{bm}% bold math
\usepackage{hyperref}% add hypertext capabilities
\usepackage{xcolor}
\newcommand{\ket}[1]{\left|{#1}\right\rangle}
\newcommand{\bra}[1]{\left\langle{#1}\right|}
\newcommand{\avg}[1]{\left\langle{#1}\right\rangle}
\newcommand{\braket}[2]{\left\langle{#1}\vert{#2}\right\rangle}
\newcommand{\sigmaz}{\hat{\sigma}_z}
\newcommand{\sigmam}{\hat{\sigma}_-}
\newcommand{\sigmap}{\hat{\sigma}_+}
\newcommand{\aop}{\hat{a}}
\newcommand{\adag}{\hat{a}^\dagger}
\newcommand{\corr}[2]{G^{(1)}\left(#1,#2\right)}

\begin{document}

\preprint{APS/123-QED}

%\title{Coherent and incoherent radiation of a transmon qubit driven by a microwave pulse}
%\title{Direct observation of the field-qubit interaction dynamics}
\title{Evolution of coherent waves driving a single artificial atom}
% Author 1
\author{A. V. \surname{Vasenin}}
 \email{vasenin.av@phystech.edu}
 \affiliation{Skolkovo Institute of Science and Technology, Nobel Street 3, 143026 Moscow, Russia}
 \affiliation{Laboratory of Artificial Quantum Systems, Moscow Institute of Physics and Technology, 141701 Dolgoprudny, Russia}

% Author 2
\author{Sh. V. \surname{Kadyrmetov}}
 \affiliation{Laboratory of Artificial Quantum Systems, Moscow Institute of Physics and Technology, 141701 Dolgoprudny, Russia}

% Author 3
\author{A. N. \surname{Bolgar}}
 \affiliation{Laboratory of Artificial Quantum Systems, Moscow Institute of Physics and Technology, 141701 Dolgoprudny, Russia}

% Author 4
\author{A. Yu. \surname{Dmitriev}}
\affiliation{Laboratory of Artificial Quantum Systems, Moscow Institute of Physics and Technology, 141701 Dolgoprudny, Russia}

% Author 5
\author{O. V. \surname{Astafiev}}
 \affiliation{Skolkovo Institute of Science and Technology, Nobel Street 3, 143026 Moscow, Russia}
 \affiliation{Laboratory of Artificial Quantum Systems, Moscow Institute of Physics and Technology, 141701 Dolgoprudny, Russia}

\date{\today}

\begin{abstract}
An electromagnetic wave propagating through a waveguide with a strongly coupled superconducting artificial two-level atom exhibits an evolving superposition with the atom. The Rabi oscillations in the atom result from a single excitation-relaxation, corresponding to photon absorption and stimulated emission from/to the field. In this study, we investigate the time-dependent behavior of the transmitted field and extract its spectra. The scattered fields are described using input-output theory. We demonstrate that the time evolution of the propagating fields, due to interaction, encapsulates all information about the atom. Additionally, we deduce the dynamics of the incoherent radiation component from the measured first-order correlation function of the field.
\end{abstract}

\maketitle

\textit{Introduction. --- }
Waveguide Quantum Electrodynamics (w-QED) involves artificial atoms that couple strongly to the propagating field \cite{Hoi2013, Gu2017, Roy2017, Kockum2020}. The first successful experiments with superconducting artificial atoms established the foundation for w-QED \cite{Astafiev2010, Hoi2011}. Subsequently, the focus shifted to more complex systems, including quantum optical effects with three-level atoms \cite{Astafiev2010a, Abdumalikov2010, HoeniglDecrinis2018} and scattering by many qubits coupled to a waveguide \cite{Loo2013, Kannan2020, Kannan2020a}. Researchers demonstrated the use of coherent and incoherent emission for inferring a qubit state \cite{Abdumalikov2011}. Other practical applications of qubit-waveguide systems include the creation of efficient single-photon sources \cite{Peng2016, Zhou2020, lu2021quantum}, and coupling two distant qubits by a waveguide to observe their entanglement \cite{Loo2013, Storz2023}. Despite reported prototypes \cite{inomata2016single, kono2018quantum, besse2018single}, robust detection of gigahertz single photons remains challenging. Nevertheless, tomographic methods utilizing linear field detectors reliably characterize microwave fields emitted by superconducting quantum systems \cite{Mallet2011quantum, eichler2011experimental}. These methods enable the resolution of peculiar properties of resonance fluorescence \cite{campagne2014observing, toyli2016resonance}, as well as the characterization of numerous non-Gaussian states of light \cite{Kudra2022robust} and non-classical multi-photon states with time-bin entanglement \cite{kurpiers2019quantum}.

Despite the aforementioned achievements, gaps still exist in experimental characterization of non-stationary resonance fluorescence of a two-level atom \cite{renaud1977nonstationary, Eberly_1980}. Researchers primarily study driven two-level systems in the stationary regime \cite{lu2021characterizing}, focusing on stationary fluorescence obtained through measurements of time-averaged spectral properties \cite{Abdumalikov2010} or stationary radiation fields \cite{lu2021propagating}. Previous attempts to observe the dynamics of the driving field were limited to spontaneous emission \cite{Sharafiev2021, Gunin2023} or very short pulses \cite{Dmitriev2017, Dmitriev2019, Vasenin2022}. However, none of these works offers a complete picture of the phenomenon, specifically a thorough time-domain characterization of a propagating pulse that interacts with the atom and then flies away. Nonetheless, such a technique is necessary for observing the properties of non-stationary resonance fluorescence.

\begin{figure}[b]
\includegraphics[width=\linewidth]{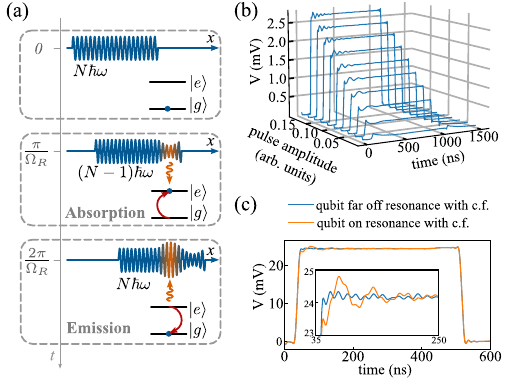}
\caption{\label{fig:basics}Stimulated absorption and emission observed in a modulated pulse shape scattered by a two-level atom. (a) Evolution of the pulse shape as it travels through a waveguide coupled to a two-level atom. (b) Voltage traces of the modulated pulse transmitted through a coplanar waveguide with a strongly-coupled transmon qubit. (c) Modulation caused by stimulated absorption and emission processes compared to an unmodulated pulse shape.}
\end{figure}

In this work we study the quantum dynamics of propagating electromagnetic waves interacting with an artificial atom strongly coupled to the field. Due to the atom-field superposition produced from this interaction, the scattered wave carries information about the atomic evolution. Figure \ref{fig:basics}(a) illustrates a gedankenexperiment, where a rectangular pulse of a coherent field $|\alpha\rangle$ with a carrier frequency matching the qubit resonance traverses an infinite waveguide. After a certain time, it causes photon absorption and excites the qubit. As a result, the pulse amplitude and power decrease. However, the continuing pulse stimulates the qubit to emit its excitation, thus increasing the pulse amplitude and power higher than their initial magnitudes. These processes repeat until the pulse ends. Decoherence causes the modulating oscillations to decay. After the pulse passes, the qubit excitation decays into the waveguide vacuum, producing an exponential tail, a phenomenon discussed in Ref. \cite{Astafiev2010, Abdumalikov2011, Sharafiev2021}.

Unlike prior methods \cite{Abdumalikov2011} that utilized repeated pulse sequences to infer qubit states from post-pulse spontaneous emission, we directly observe stimulated photon dynamics between the qubit and the field during pulse drive, extracting Rabi oscillations, qubit population, and the first-order correlation function. The first-order correlation function and instantaneous power density spectra reveal the dynamics of incoherent radiation and confirm the quantum nature of the observed phenomenon. Notably, as the averaged field oscillations diminish, the field correlation function retains its oscillations, highlighting the interplay between coherent and incoherent fluorescence in a non-stationary regime.

\textit{Theory. ---}
The model Hamiltonian for our atom-waveguide system features a two-level atom symmetrically coupled to the left- and right-moving free field modes of a one-dimensional waveguide:
\begin{multline}
    \hat{H}=-\frac{\hbar\omega_q}{2}\sigmaz
    +\sum_{k=l,r}\Biggl[\int d\omega\hbar\omega\adag_{k,\omega}\aop_{k,\omega} \\
    + \hbar g \int d\omega\left(\aop_{k,\omega}\sigmap + \adag_{k,\omega}\sigmam\right)\biggr],
\end{multline}
where $\hbar\omega_q$ denotes the qubit energy splitting, $l$ and $r$ indices represent left and right modes, and $g$ is the atomic coupling constant to the waveguide field modes.

Experimentally, the right-moving mode field is measured. Per the input-output theory for a tightly coupled atom-waveguide system \cite{Gardiner1985,Fan2010,Peropadre2013,Lalumiere2013,Lindkvist2014} in the interaction picture, the field at the right waveguide output $\aop_o(t)$ corresponds to the left input field $\aop_i(t)$ as:
\begin{equation}\label{eq:input-output}
\aop_{o}(t)=\aop_{i}(t)+i\sqrt{\frac{\Gamma_1}{2}}\sigmam(t).
\end{equation}

We use IQ-mixers which downconvert the signal and produce two field quadratures: $\hat{V}_I(t)=V_0(\aop_o(t)+\adag_o(t))/2$ and $\hat{V}_Q(t)=-iV_0(\aop_o(t)-\adag_o(t))/2$. Here, $V_0=\sqrt{G\hbar\omega Z_0}$, where $G$ represents the amplification coefficient and $Z_0$ is the waveguide impedance. Together, they define the full output field $\hat{V}_o(t)=V_I(t)+iV_Q(t)=V_0\aop_o(t)$.

For a coherent rectangular pulse driving an atom, the atom state displays Rabi oscillations at frequency $\Omega_R(t) = \sqrt{2\Gamma_1}\alpha(t)$. These oscillations, notable during the initial pumping $0 \leq t \lesssim 1/\Gamma_1$, manifest in the averaged output voltage trace:
\begin{equation}
\avg{\hat{V}_o(t)} = V_0\left(\alpha(t)+i\sqrt{\Gamma_1/2}\avg{\sigmam(t)}\right).
\end{equation}

From equation (\ref{eq:input-output}), it can be derived that the energy absorbed from the transmitted pulse in the first half of Rabi period nearly matches the single-photon energy. The same applies for the energy emitted in the subsequent half cycle. Conversely, energy reflected in the left direction is minimal. Analytical calculations are detailed in \cite{SuppMat}.

While \(\avg{\hat{V}_o(t)}\) reveals the coherent part of radiation, the non-stationary first-order correlation function exposes the incoherent component of the field:
\begin{multline}
    \label{eq:g1-incoh}
    \corr{t_1}{t_2}=V_0^2\big(\avg{\adag_o(t_1)\aop_o(t_2)}\\-\avg{\adag_o(t_1)}\avg{\aop_o(t_2)}\big).
\end{multline}

For a classical coherent wave in a transmission line, without qubit interaction, the average field is \(\avg{\hat{V}_o(t)}=V_0\alpha(t)\) and \(\corr{t_1}{t_2}=0\), with \(\vert\alpha(t)\vert^2\) being the photon rate. With atom interaction, \(\corr{t_1}{t_2}\) becomes:
\begin{multline}\label{eq:qubit-g1}
\corr{t_1}{t_2} = \frac{V_0^2\Gamma_1}{2}\big(\avg{\hat{\sigma}_+(t_1)\hat{\sigma}_-(t_2)}\\
-\avg{\hat{\sigma}_+(t_1)}\avg{\hat{\sigma}_-(t_2)}\big).
\end{multline}

The first-order correlation function highlights the \(\aop\) operator property to maintain a coherent state \(\aop\ket{\alpha}=\alpha\ket{\alpha}\) and \(\adag\) to disrupt coherence \cite{Agarwal1991}. When evolution operator $\hat{U}(\delta t)\approx 1-ig\delta t(\aop(t)\sigmap+\adag(t)\sigmam)$, assuming $\alpha(t)g\delta t\ll 1$, acts on the atom-field wavefunction $\ket{\Psi_q(t),\alpha(t)}$, it preserves the coherent state of the field if 
\begin{equation}
    \hat{U}(\delta t) \ket{g}\ket{\alpha(t)} \approx (\ket{g} - ig\alpha(t)\delta t \ket{e}) \ket{\alpha(t)}
\end{equation} and spoils it if
\begin{equation}
    \hat{U}(\delta t) \ket{e}\ket{\alpha(t)} \approx (\ket{e}\ket{\alpha(t)} - ig\delta t\ket{g} \adag(t) \ket{\alpha(t)}).
\end{equation}
Therefore, it is expected that $\corr{t}{t+\tau}\approx0$ at $t=\frac{2\pi n}{\Omega_R}$ and $\corr{t}{t+\tau}\neq 0$ at $t=\frac{\pi(2 n - 1)}{\Omega_R}$, as confirmed by our experiments.

\textit{Results and discussion. --- }
We first measure the averaged voltage trace of a pulse on resonance with the qubit. Fig. \ref{fig:basics}(b) displays oscillations near the front of the pulse, which increase in frequency with pulse amplitude while maintaining their amplitude consistent. In Fig. \ref{fig:basics}(c), comparing the pulse shape when qubit is on-resonance and far-detuned reveals that modulation begins with a decrease in pulse amplitude, likely signifying photon absorption. To further investigate this modulation, we subtract the two measurements.

\begin{figure}[b]
\includegraphics[width=\linewidth]{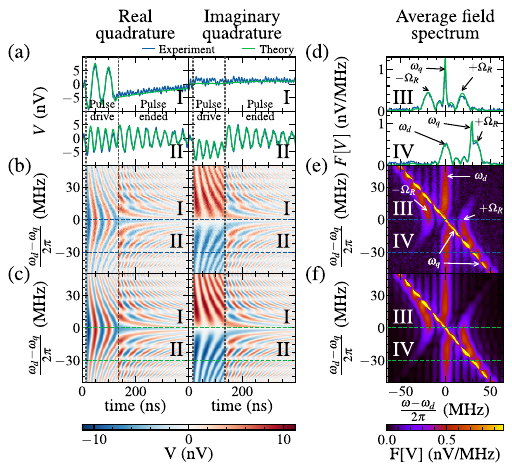}
\caption{\label{fig:oscillations_vs_detuning}Radiation of a qubit as a function of detuning of a pulse carrier frequency ($\omega_d$) from the qubit transition frequency ($\omega_q$). The radiation is defined as the difference between a modulated pulse and an unperturbed pulse. The pulse is rectangular with 120 ns length. (a) Single traces of radiation for resonance and slightly off-resonance (-30 MHz) cases. (b) Experimental traces plotted in color as a function of detuning. (c) Results of a simulation for radiation traces. (d) Magnitude spectra of single traces taken from (a). Arrows indicate peaks corresponding to Rabi, qubit and drive frequencies. (e) Experimental spectra plotted in color as a function of detuning. (f) Simulated spectra plotted in color.}
\end{figure}

Fig. \ref{fig:oscillations_vs_detuning}(a) shows such differences for a resonant case (I) and a slightly off-resonant case (II) when the carrier frequency deviated from the qubit frequency. We measure and digitally rotate two quadratures of the field so that the real quadrature is in phase with the pulse, and the imaginary quadrature has an orthogonal phase. Resonant traces display oscillations during the pulse drive and an exponential decay post-pulse. Their magnitude spectra are in Fig. \ref{fig:oscillations_vs_detuning}(d), with the horizontal axis centered at the pulse carrier frequency. A distinct peak in the spectra corresponds to post-pulse oscillations, flanked by two broader peaks. Simulation of $\avg{\sigmam(t)}$ identifies these traces as Rabi oscillations. The sharp peak matches the qubit frequency, while the wide peaks represent positive and negative Rabi frequencies in the resonant trace. For the off-resonant trace, the right peak aligns with the positive Rabi frequency, and the left peak is at 0 Hz, indicating a constant voltage offset during pulse drive.

Traces were scanned against the detuning $\delta=\omega_d-\omega_q$ from the qubit resonance. Experimental results and simulations are in Fig. \ref{fig:oscillations_vs_detuning}(b-c), with magnitude spectra in Fig. \ref{fig:oscillations_vs_detuning}(e-f). Fitting data yielded qubit parameters: qubit frequency $f_q=4.835~\text{GHz}$, Rabi frequency $\Omega_R/(2\pi)=19.8~\text{MHz}$, decay rate $\Gamma_1/(2\pi)=0.9~\text{MHz}$, and dephasing rate $\gamma/(2\pi)=0.6~\text{MHz}$. Measurements were performed not at the qubit sweet-spot because of a spurious TLS, hence large dephasing. Fig. \ref{fig:oscillations_vs_detuning}(b-c) reveals two regions: the first is the qubit radiation during the microwave drive yielding three spectral peaks at $\omega_d-\Omega$, $\omega_d$, and $\omega_d+\Omega$ where $\Omega=\sqrt{\Omega_R^2+\delta^2}$. Their width is mostly defined by pulse length. The second is the spontaneous emission \cite{Sharafiev2021} post-pulse, which yields a peak limited by $\Gamma_2$, consistent at the qubit frequency as the carrier shifts. With known Rabi frequency and decay rate, voltages were recalibrated to atom-radiated values on the chip \cite{HoeniglDecrinis2020}.

Solving the evolution of $\avg{\sigmam(t)}$ in the dressed basis \cite{CohenTannoudji1998} explains asymmetry of side-band amplitudes at $\omega_d\pm\Omega$ when $\delta\neq0$. According to this approach, when a coherent state populates one of the waveguide modes, levels $\ket{g,n+1}$ and $\ket{e,n}$ hybridize to new basis states $\ket{+n}$ and $\ket{-n}$ with energy splitting $E_{\ket{+n}}-E_{\ket{-n}} =\hbar\Omega$.
\begin{gather}
    \ket{+n} = \sqrt{\frac{\Omega+\delta}{2\Omega}}\ket{g, n+1}
                + \sqrt{\frac{\Omega-\delta}{2\Omega}} \ket{e,n}\\
    \ket{-n} = \sqrt{\frac{\Omega-\delta}{2\Omega}} \ket{g, n+1} 
                - \sqrt{\frac{\Omega+\delta}{2\Omega}} \ket{e,n}
\end{gather}
Amplitudes of the peaks at $\omega_d-\Omega$ depend on detuning as
\begin{multline}
    \braket{g,n}{+(n-1)}\bra{+(n-1)}\sigmam\ket{-n}\cdot \\
    \cdot\braket{-n}{g,n+1}=-\frac{\Omega_R(\Omega+\delta)}{4\Omega^2}
\end{multline} and at $\omega_d+\Omega$ as 
\begin{multline}
    \braket{g,n}{-(n-1)}\bra{-(n-1)}\sigmam\ket{+n}\cdot \\
    \cdot\braket{+n}{g,n+1}=\frac{\Omega_R(\Omega-\delta)}{4\Omega^2}.
\end{multline}

\begin{figure}[t]
\includegraphics[width=\linewidth]{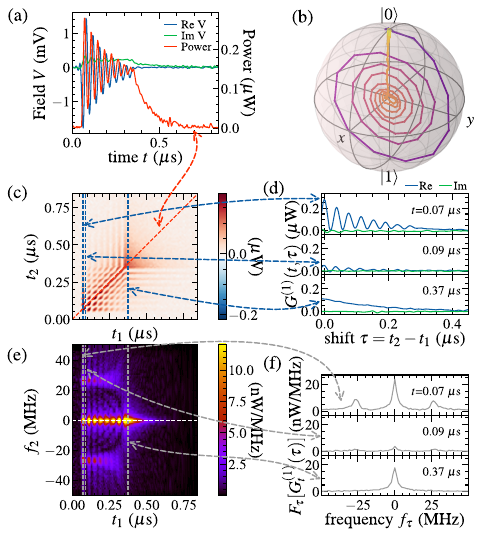}
\caption{\label{fig:first-order-correlation-functions}Measurement of the second-order momenta of the field in the resonant case. (a) Traces of field and qubit population as power. (b) Qubit state evolution extracted from traces in (a). (c) Real part of the first-order correlation function. Its diagonal trace is the population trace in (a). Vertical slices are shown in (d). (e) Fourier spectra taken along $\tau=t_2=t_1$ for the first-order correlation function. Vertical slices from (e) are shown in (f).}
\end{figure}

The first-order correlation function $\corr{t_1}{t_2}$ for a finite-length pulse was measured using equation (\ref{eq:g1-incoh}) and methods from \cite{Silva2010,Zhou2020}. Fig. \ref{fig:first-order-correlation-functions}(c) shows the real part of function (\ref{eq:qubit-g1}) for the resonant case, while Fig. \ref{fig:first-order-correlation-functions}(d) displays 2D correlation function slices, indicating the imaginary part is near zero. The central square in Fig. \ref{fig:first-order-correlation-functions}(c) highlights correlations during the pulse drive, with primarily real positive values in both resonant and non-resonant scenarios. Another square along the diagonal indicates post-pulse correlations, real and positive only in resonance. The rectangles flanking the diagonal show voltage correlations between pulse and post-pulse times. The function's real, positive nature during the pulse underscores the quantum nature of the effect, since classical coherent oscillating fields cannot exhibit the same property. Additionally, the averaged field, combined with the diagonal trace of the correlation function, provides a complete tomography of the qubit state evolution.

To analyze the measured $\corr{t_1}{t_2}$, a Fourier transform is performed as
\begin{equation}\label{eq:ipsd}
IPSD=\left\vert \int_{0}^{+\infty}\corr{t}{t+\tau}e^{-i\omega\tau} d\tau\right\vert,
\end{equation}
with $\tau=t_2-t_1 > 0$ representing the time difference between two points. The IPSD (instant power spectral density) reveals three distinct peaks (Fig. \ref{fig:first-order-correlation-functions}(e)) during the pulse drive, with sidebands at Rabi frequencies. Post-pulse, only the central peak remains, related to spontaneous emission.

\begin{figure}[t]
\includegraphics[width=\linewidth]{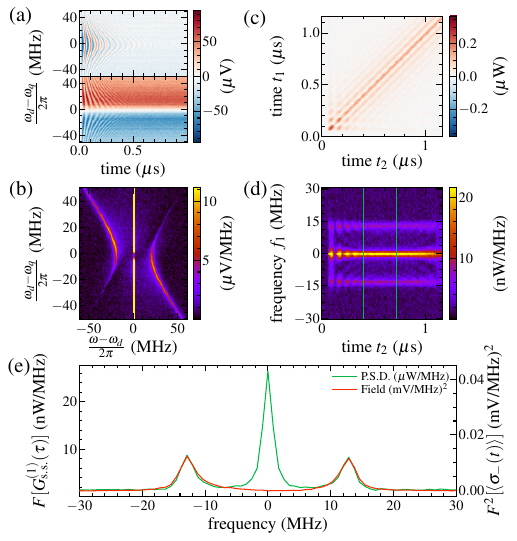}
\caption{\label{fig:long-pulse}Properties of qubit radiation during a long pulse with a large amplitude. (a)-(b) Dynamics of the coherent field and its spectra as a function of the frequency difference between the qubit resonance and the carrier frequency. (c)-(d)  Non-stationary two-time first-order correlation function and instant P.S.D. for the pulse on resonance. (e) Comparison of the squared spectrum of the average field with the power spectral density in the steady state, which exhibits a Mollow triplet. To smooth the P.S.D. we averaged it over the stationary period between the two green lines in the subfigure (d).}
\end{figure}

A long pulse measurement (Fig. \ref{fig:long-pulse}) enhances spectral resolution. Stimulated emission peaks in the field spectra become narrower, constrained by $(\Gamma_1+\Gamma_2)/2$. At $\omega_d \neq \omega_q$, a sharp peak appears above a broader one with $\Gamma_2/2$ width. The free decay spectral line is absent in Fig. \ref{fig:long-pulse}(b) because we have cut voltage traces at the end of the pulse. Fig. \ref{fig:long-pulse}(c) depicts the real part of the first-order correlation function for the in-resonance case, covering the initial non-stationary period and the steady period $t > 0.5$ $\mu$s, where qubit state oscillations have faded. In this steady region, the correlation function oscillates with the time difference $\tau=t_2-t_1$. Fourier transforming along $\tau$ and averaging within the steady region (between green lines in Fig. \ref{fig:long-pulse}(d)) produces a power spectral density (PSD) resembling a Mollow triplet \cite{Mollow1969, Astafiev2010}. Comparing this with the squared spectrum of the averaged field in resonance shows that the sidebands align while the central peak is absent due to its incoherent nature.

As discussed in the theory section, our results (Fig. \ref{fig:first-order-correlation-functions}(c-f)) demonstrate that $\corr{t}{t+\tau}$ approaches zero for any $\tau$ at times $t$ between two adjacent Rabi periods, while it is non-zero at other time points. For instance, all peaks in IPSD (Fig. \ref{fig:first-order-correlation-functions}(e) and \ref{fig:long-pulse}(d)) exhibit oscillations with the Rabi frequency along the $t_1$ axis. In Fig. \ref{fig:long-pulse}(c, d), it is also observed that decoherence gradually diminishes the initially coherent Rabi oscillations for a very long pulse, as time progresses beyond $t > 1/\Gamma_1$ where $\corr{t}{t+\tau}$ no longer turns to zero.

\textit{Conclusion. ---}
In our experiments, we examined a coherent rectangular pulse evolution as it traversed a 1D waveguide with a strongly coupled superconducting artificial atom. We demonstrated that the artificial atom modulates the pulse field with Rabi oscillations. Our findings align closely with simulations based on the input-output theory expression (\ref{eq:input-output}). The field spectra were interpreted in terms of transitions between dressed qubit levels. Measuring the non-stationary first-order correlation function $\corr{t_1}{t_2}$ provided insights into emitted incoherent radiation dynamics and highlighted the $\adag$-operator's inherent property to disrupt a coherent state $\ket{\alpha}$.

Earlier quantum optical experiments mainly observed the exponential tail post-pulse, which encapsulates the energy of a photon over an extended decay time. Yet, within the pulse, the qubit typically absorbs and emits a photon. Given the swift Rabi oscillations, this photon energy is confined to a briefer span, resulting in power oscillations of greater amplitude than the exponential tail. Thus, beyond quantum optics, qubit state measurements also can benefit from measuring modulation of a driving pulse after interaction with the atom.

% Acknowledgments
\begin{acknowledgments}
We wish to acknowledge the support of Russian Science Foundation (grant N 21-42-00025). This work was performed using equipment of MIPT Shared Facilities Center.
\end{acknowledgments}

\bibliography{bibliography}

\end{document}

\title{Supplemental Material for\\Evolution of coherent waves driving a single artificial atom}

\date{\today}
\maketitle

\tableofcontents

\section{Measurement details}
\begin{figure}
    \centering
    \includegraphics[width=0.75\linewidth]{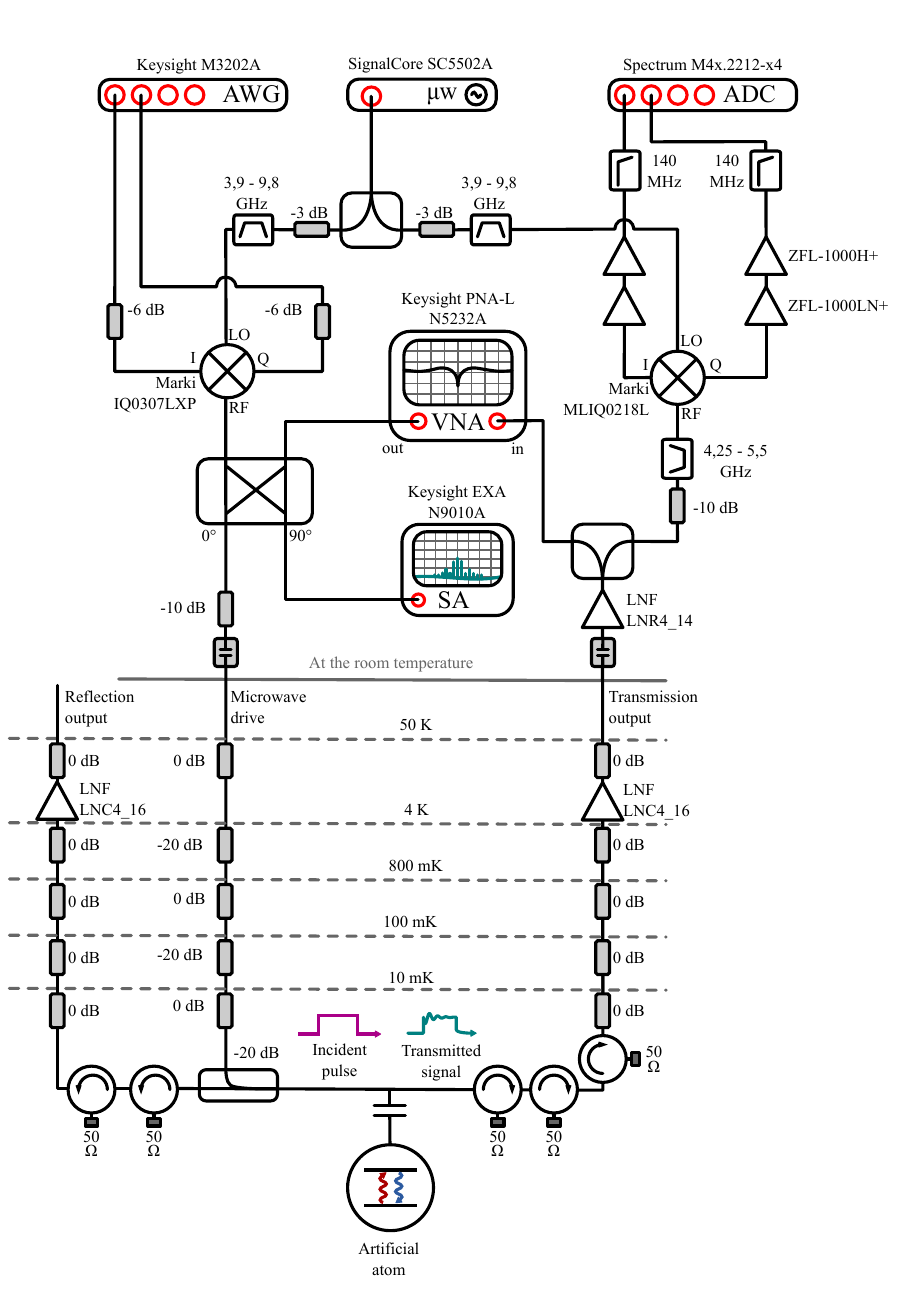}
    \caption{Measurement setup}
    \label{fig:measument-setup}
\end{figure}
A complete scheme of our measurement setup is shown in Fig. \ref{fig:measument-setup}. Our experiments utilize a transmon qubit as a two-level artificial atom strongly capacitively coupled to a coplanar waveguide (CPW). We cool the chip with a qubit down to 10-15 mK using a dilution refrigerator. The output line contains two cryogenic isolators at the mixing chamber plate and a HEMT amplifier at the 4K plate. We fix the carrier frequency of the pulses on resonance with the transmon upper sweet-spot at 5 GHz. By tuning the qubit to its lower sweet-spot at 3.5 GHz with an external magnetic field, we can disable the interaction between the pulses and the qubit. We measure the pulses transmitted through the sample using a heterodyne setup consisting of two IQ-mixers, a microwave generator, an arbitrary waveform generator (AWG), and a digitizer. As the qubit emits photons within time $T_1=\Gamma_1^{-1}=114$~ns, we set the repetition to $1-2 \mu$s.

To observe modulation of transmitted pulse shape, we use an AWG to prepare a cosine-tapered pulse at an intermediate frequency of 100 MHz, which is then converted up to a 5 GHz carrier frequency in resonance with the upper sweet-spot of the qubit. After transmission through the sample, the pulse is amplified and measured using an ADC that can average over multiple trigger events. The measured voltage trace is down-converted to 0 Hz and filtered to remove frequencies above 50 MHz, yielding the pulse shape.

The signal (spontaneously emitted photon) to noise ratio for a single measurement is calculated to be $\hbar\omega/(k_B T_\text{noise})\approx-40~\text{dB}$. We compute average traces and correlation functions with a real-time GPU processor from the measured data. The voltage traces are averaged over 17 million measurements, taking around 17 seconds in total. Correlation functions require at least a billion averaging iterations and take not less than 33 minutes to measure.

\section{Qubit parameters}
Fig. 2 from the main text was fitted with the following parameters: qubit frequency $f_q=4.835~\text{GHz}$, Rabi frequency $\Omega_R/(2\pi)=19.8~\text{MHz}$, decay rate $\Gamma_1/(2\pi)=0.9~\text{MHz}$, dephasing rate $\gamma/(2\pi)=0.6~\text{MHz}$. Zero-temperature model is assumed because non-zero temperate model is overfitting. The pulse was cosine-tapered with a duration of 120 ns. The tapered part takes 2 \% of the whole duration. An estimated photon rate of the rectangular pulse was $\nu=\Omega_R^2/(2\Gamma_1)=1.4~\text{photons/ns}$.

\section{Voltage amplitude of the qubit radiation going to the right}
In this sections, we will show that the constant in front of $\sigmam(t)$ in the radiation field is $e^{i\varphi}\sqrt{\frac{\Gamma_1}{2}}$.
Imagine an atom emitting a photon into a vacuum of an infinite one-dimensional bidirectional waveguide. The population of its excited state decays exponentially as $P(t)=e^{-\Gamma_1t}$. The integral of $P(t)$ is
\begin{equation}
    \int\limits_0^{+\infty} P(t)dt=\int\limits_0^{+\infty} e^{-\Gamma_1t}dt=\frac{1}{\Gamma_1}\int\limits_0^{+\infty} e^{-x}dx=\frac{1}{\Gamma_1}.\label{eq:population}
\end{equation}
The field emitted by an atom into the right direction can be defined by $\hat{V}_\text{out}(t)=\varkappa\sigmam(t)$. Then the power of the emission is $W(t)=\frac{\vert\varkappa\vert^2}{Z_0}\avg{\sigmap(t)\sigmam(t)}=\vert\varkappa\vert^2P(t)$.

Because the atom is coupled equally to the left-moving and right-moving modes in the infinite waveguide and emits a photon with energy $\hbar\omega_q$, on average only half a photon gets measured in the power from the right (or left) end of a waveguide. This gives
\begin{equation}
    \frac{\hbar\omega_q}{2}=\int\limits_0^{+\infty} W(t)dt=\frac{\vert\varkappa\vert^2}{Z_0}\int\limits_0^{+\infty} P(t)dt=\frac{\vert\varkappa\vert^2}{Z_0\Gamma_1}.\label{eq:half-photon}
\end{equation}

From here it follows that
\begin{equation}
    \varkappa=\sqrt{\frac{\hbar\omega_q Z_0\Gamma_1}{2}}e^{i\varphi}\label{eq:kappa}.
\end{equation}

Taking into consideration, that the emitted field should remain congruent when the external pulse stops driving the qubit, we conclude that $\varkappa=\mathrm{const}$.

If the coherent pulse $V_p(t)=\sqrt{\hbar\omega Z_0}\alpha(t)$ drives an atom (here $\vert\alpha(t)\vert^2$ is the photon rate of the incoming radiation), then the resulting field at the right output of the waveguide must be
\begin{equation}\label{eq:voltage-with-pulse}
    V_\text{out}=V_p(t)+V_q(t)=\sqrt{\hbar\omega Z_0}\left(\alpha(t)+\sqrt{\frac{\Gamma_1}{2}}e^{i\varphi}\avg{\sigmam(t)}\right).
\end{equation}

The phase $\varphi$ depends on the choice of the interaction Hamiltonian. If one chooses $\hat{H}_\text{int}=g(\adag\sigmam+\aop\sigmap)$, the phase would be $\varphi=\frac{\pi}{2}$ and the measured field will be proportional to $\alpha(t)+i\sqrt{\frac{\Gamma_1}{2}}\avg{\sigmam}$. Otherwise, if $\hat{H}_\text{int}=ig(\adag\sigmam-\aop\sigmap)$ was chosen, then the phase would be $\varphi=0$ and the output field will be proportional to $\alpha(t)+\sqrt{\frac{\Gamma_1}{2}}\avg{\sigmam(t)}$.

Since $\frac{\Omega_R}{2}=\sqrt{\frac{\Gamma_1}{2}}\alpha$, we get that
\begin{equation}\label{eq:voltage-with-pulse}
    V_\text{out}(t)=V_p(t)+V_q(t)=\sqrt{\hbar\omega Z_0}\left(\frac{\Omega_R(t)}{\sqrt{2\Gamma_1}}+\sqrt{\frac{\Gamma_1}{2}}e^{i\varphi}\avg{\sigmam(t)}\right)=\sqrt{\frac{\hbar\omega Z_0\Gamma_1}{2}}\left(\frac{\Omega_R(t)}{\Gamma_1}+e^{i\varphi}\avg{\sigmam(t)}\right).
\end{equation}

\section{Number of photons in a coherent pulse}
There are two methods to find the number of photons in a rectangular pulse. The first one follows directly from the previous sections uses only the value of $\Gamma_1$. The second method requires two values $\Gamma_1$ and $\Omega_R$.

\subsection{Number of photons from the comparison of the radiation amplitude with the incoming pulse amplitude}
Taking into account the pulse amplitude $V_p$ and oscillations amplitude $V_q$, one should first calculate the ratio $\frac{V_p}{V_q}$.

From the equation (\ref{eq:voltage-with-pulse}), it follows that
\begin{equation}
    \frac{V_p}{V_q}=\frac{2\sqrt{2}\vert\alpha\vert}{\sqrt{\Gamma_1}}.
\end{equation}

Thus one can compute the photons rate of the pulse as
\begin{equation}
    \nu=\vert\alpha\vert^2=\frac{\Gamma_1}{8}\left(\frac{V_p}{V_q}\right)^2.
\end{equation}

\subsection{Extracting number of photons from the Rabi frequency}
The second method relies on the formula $\frac{\Omega_R}{2}=g\alpha$. Where $g$ is the coupling constant in the qubit-waveguide interaction Hamiltonian $H_\text{int}=g\left(\adag(t)\sigmam(t)+\aop(t)\sigmap(t)\right)$. In case of a qubit in the middle of the one-dimensional bidirectional waveguide, $g=\sqrt{\frac{\Gamma_1}{2}}$. Therefore, the photon rate of a pulse is
\begin{equation}
    \nu=\frac{\Omega_R^2}{2\Gamma_1}.
\end{equation}

\section{Energy absorbed from the pulse in a half of a Rabi period}
For our atom-waveguide system, the field propagating to the right is 
\begin{equation}
    \aop_o^{(r)} = \aop_i^{(r)}+i\sqrt{\frac{\Gamma_1}{2}}\sigmam(t)
\end{equation}
and the field propagating to the left is
\begin{equation}
    \aop_o^{(l)} = i\sqrt{\frac{\Gamma_1}{2}}\sigmam(t).
\end{equation}

With the input field $\ket{\alpha(t)}$, the average power going right will be
\begin{equation}
    P^{(r)}_o(t)=\vert\alpha(t)\vert^2-\alpha(t)\sqrt{\frac{\Gamma_1}{2}}\avg{\sigmay(t)}+\frac{\Gamma_1}{2}\frac{1-\avg{\sigmaz(t)}}{2}
\end{equation}
and the average power going to the left will be
\begin{equation}
    P^{(l)}_o(t)=\frac{\Gamma_1}{2}\frac{1-\avg{\sigmaz(t)}}{2}
\end{equation}

From now on, we assume the decay rate $\Gamma_1$ to be much smaller than Rabi oscillations frequency $\Omega_R(t)=\sqrt{2\Gamma_1}\alpha(t)$ and the qubit is driven in resonance, so that the qubit evolution is approximately the following
\begin{gather}
    \avg{\sigmax(t)}=0,\\
    \avg{\sigmay(t)}=\sin(\Omega_Rt),\\
    \avg{\sigmaz(t)}=\cos(\Omega_Rt).
\end{gather}

We get the energy of the transmitted part of pulse after the first half of Rabi period by integrating the right-propagating power from 0 to $\pi/\Omega_R$
\begin{equation}
    E^{(r)}\left(\frac{\pi}{\Omega_R}\right)=\int\limits_0^{\pi/\Omega_R}P_o^{(r)}(t)dt=N-1+\frac{\pi\Gamma_1}{4\Omega_R},\label{eq:right-energy}
\end{equation}
where $N=\frac{\pi\vert\alpha\vert^2}{\Omega_R}$ is the average number of photons that was in this part of pulse before interaction with a qubit. The last term in (\ref{eq:right-energy}) is very small compared to 1 under our assumption $\Gamma_1\ll\Omega_R$ and also equals to $E^{(l)}\left(\frac{\pi}{\Omega_R}\right)$.

Equation (\ref{eq:right-energy}) shows that the energy of a single photon was subtracted from the initial pulse energy after the first half of Rabi period. It can be shown, that during the next half of Rabi period, the energy of a single photon will be returned to the pulse.